\documentclass[aps,pre,twocolumn,showkeys,floatfix,groupedaddress,longbibliography]{revtex4-1}

\usepackage{amsmath}
\usepackage{graphicx}
\usepackage[caption=false]{subfig}
\usepackage{verbatim}

\graphicspath{{Plots/}}
\begin{document}

%\preprint{}

\title{A Fluctuating Lattice Boltzmann Method for the Diffusion Equation}


\author{Alexander J. Wagner}
\email[]{alexander.wagner@ndsu.edu}
\homepage[]{www.ndsu.edu/pubweb/~carswagn}
\author{Kyle Strand}
%\thanks{}
\affiliation{Department of Physics, North Dakota State University, Fargo, ND 58108}


\date{\today}

\begin{abstract}
We derive a fluctuating lattice Boltzmann method for the diffusion equation. The derivation removes several shortcomings of previous derivations for fluctuating lattice Boltzmann methods for hydrodynamic systems. The comparative simplicity of this diffusive system highlights the basic features of this first exact derivation of a fluctuating lattice Boltzmann method.
\end{abstract}

%\pacs{}

\maketitle

\section{Introduction}
This paper introduces a new lattice Boltzmann method for the fluctuating diffusion equation. Much of this work was inspired by earlier work on fluctuating hydrodynamics which resulted in very decent, but ultimately never quite exact discrete representations. This is why we focus here on the arguably simplest lattice Boltzmann method to extend it to its fluctuating counterpart. As we show below this derivation allows us to remove certain inexact assumptions of earlier methods leading to a surprisingly exact and robust fluctuating method. Even the limit of low density, where few particles reside on each lattice site on average, is well behaved. This is unexpected since in a continuous method the noise starts to fully dominate the dynamics of the system at such low densities, and for hydrodynamics fluctuating lattice Boltzmann methods this limit is ill behaved \cite{kaehlerUnpub}.

Let us step back for a moment to consider the interplay of fluctuations and continuous methods. Materials are composed of discrete particles that follow the deterministic evolution of Newton's second law (or more fundamentally the laws of quantum mechanics). The collective evolution of many of these particles evolves through rules that can be discovered through kinetic theory. In many cases there is a fast dynamic due to the local collisions between particles and a slow dynamics of locally conserved quantities that are unchanged by the collisions. Due to the chaotic nature of the collisions the evolution of the locally conserved quantities has a deterministic part (i.e. the hydrodynamic equation) and a non-deterministic part that can often be well described as an instantaneous random component. The amount of the random component will depend on the amount of coarse-graining, and it is often assumed that for macroscopic systems these fluctuations will be averaged out, leaving the fully deterministic hydrodynamic equations.

The reason for the fluctuations lies in the discrete nature of our world. It is often argued that the explanation of random motion of small particles, known as Brownian motion, in terms of individual collisions with the molecules of a fluid by Einstein\cite{Renn} was the clinching argument in favor of the particulate nature of matter. But a continuous description of nature has jettisoned this fundamentally discrete nature of matter, and therefore fluctuations have to be artificially re-introduced. This is achieved through the Langevin method, where the idea is to introduce random forces such that the predicted level of fluctuation is recovered in equilibrium \cite{Langevin}. These fluctuations obey a fluctuation-dissipation theorem where the irreversible dissipative parts of the evolution are balanced by their fluctuating counterpart in equilibrium.

For the Navier-Stokes equation the fluctuating equivalent was derived by Landau and Lifshitz \cite{LandauLifshitz}. It consists of complementing the dissipative viscous stress tensor with a fluctuating stress tensor. The first fluctuating lattice Boltzmann method, introduced by Ladd in 1993 \cite{LaddPRL,LaddJFM}, used this result and introduced a fluctuating stress tensor into the lattice Boltzmann method. However, this only gives correct results in the hydrodynamic limit, i.e. for small wave numbers in a Fourier representation.

Earlier work around 1970 of Fox and Uhlenbeck \cite{Fox} as well as Bixon and Zwanzig \cite{Bixon} generalized the linear Boltzmann equation into a Langevin equation by introducing a fluctuating collision operator. Crucially this fluctuating collision operator not only introduces a fluctuating stress tensor but additional fluctuating higher order modes. The idea of using this as a basis for the development of fluctuating lattice Boltzmann even predates the approach by Ladd, and was outlined in a conference proceeding by Dufty and Ernst in 1993 \cite{dufty} around the same time as Ladd introduced his method. Overcoming the difficulties with this approach was not achieved until 2005 when Adhikari et al. \cite{adhikari} presented a practical implementation. The key improvement introduced here was to include fluctuations not only for the stress tensor, but for all non-conserved modes that the collision operator relaxes. This significantly improved the behavior of the fluctuating lattice Boltzmann method for larger wave numbers. This approach was then re-derived by Duenweg from considerations of a discrete lattice gas \cite{duenweg}. However, there were two inconsistencies in this approach, both inherited from trying to apply the Fox-Uhlenbeck approach for a linear Boltzmann equation to a BGK-collision operator. In the linear approach, the collision is assumed to bring the local distribution function closer to the global equilibrium distribution (with special projections that still ensure local conservation of conserved quantities) whereas the BGK approach consists of a relaxation towards a local equilibrium distribution, corresponding to the local conserved quantities. A key difference is that the hydrodynamic limit of the linear Boltzmann equation does not recover the non-linear term $\nabla(\rho \mathbf{uu})$ in the Navier-Stokes equation, whereas the BGK collision operator does.

The first inconsistency occurred because in the derivation the local equilibrium distribution was replaced by the global equilibrium distribution at a crucial juncture, a problem that has been addressed by Kaehler et al. \cite{kaehlerPRE}. The second inconsistency occurred because it was assumed that the fluctuations of the local equilibrium distribution around the global equilibrium distribution could be neglected. At its heart, the current paper deals with avoiding this incorrect assumption, albeit in the simpler case of a lattice Boltzmann method for diffusion, rather than hydrodynamics. An extension to the hydrodynamic case appears also possible, but is more technically involved and will be reserved for a later publication.

In this paper we first introduce the fluctuating lattice Boltzmann method, and derive the form of the fluctuating collision operator for the diffusion equation. As usual for fluctuating lattice Boltzmann methods since Adhikari et al. \cite{adhikari} this results in a multi-relaxation-time method \cite{higuera}. While this method is distinct, lattice Boltzmann methods for diffusion phenomena were derived by Wolf-Gladrow \cite{wolf}, and can be obtained in the limit of no flow from the more popular multi-component lattice Boltzmann methods \cite{shan,orlandini}. A multi-relaxation time method related to the current method (albeit with a slightly different collision matrix) can be found in Li et al. \cite{LiPRE2014} and Le \cite{LePRE2015} (all of these methods, however, exclude fluctuations). In the following section we show that this method does indeed recover the fluctuating diffusion equation in the hydrodynamic limit and the last section shows a number of verifications of our new method, and its fluctuating qualities. We close by giving an outline of future research directions, particularly the impact of this work on fluctuating lattice Boltzmann methods for multi-phase systems, a problem that has been previously addressed by Gross et al. \cite{Gross}, Thampi et al. \cite{Thampi} and by Belardinelli et al. \cite{Belardinelli}.

    \section{Lattice Boltzmann Method}
    The lattice Boltzman equation is an evolution equation for densities $f_i$ defined on a regular lattice associated with discrete velocities $v_i$. These densities move in the direction of their associated velocities, and the velocities are typically chosen so that the resulting position is again a lattice position. Densities that collect at one lattice site then undergo local collisions. This can be formally written as
    \begin{equation}
      f_i(\mathbf{x}+\mathbf{v}_i, t+1) = f_i(\mathbf{x},t)+\Omega_i.
      \label{LB}
    \end{equation}
    Following Qian et al. \cite{Qian} one typically makes the assumption that the collision is well approximated by a BGK approximation as simply approaching a local equilibrium distribution $f_i^0$ which is a function of the locally conserved quantities. To include fluctuations we must consider not only the dissipative relaxation towards the local equilibrium, but also an additional fluctuating term $\xi_i$. We can then write the full collision term as
    \begin{equation}
      \Omega_i = \sum_j \Lambda_{ij} [f_j^0-f_j(x,t)]+\xi_i
      \label{collision}
    \end{equation}
    where $\Lambda_{ij}$ is a collision matrix.
    The purpose of this paper is to derive a convenient (and exact) form of the noise terms $\xi_i$ as a function of the collision matrix $\Lambda_{ij}$.

    In this paper we will focus on the simplest case of an ideal gas. For an ideal gas of discrete lattice particles associated with the discrete velocities $v_i$, we expect their occupation numbers $n_i$ to follow a binomial distribution. For a large enough system this is well approximated by the Poisson distribution
    \begin{equation}
      P(n_i) = \frac{\exp(-n_i^{eq})(n_i^{eq})^{n_i}}{n_i!}.
      \label{Poisson}
    \end{equation}
    Here we have defined the a priori unknown global equilibrium occupation numbers as
    \begin{equation}
      n_i^{eq} = <n_i>,
    \end{equation}
    where $<\cdots>$ denotes an equilibrium average. These fluctuations are independent for different velocities and different lattice sites. We show in the appendix, eqn. (\ref{Poisson2}) that the second moment is given by
    \begin{equation}
      <n_i(\mathbf{x},t)n_j(\mathbf{x}',t)>=n_i^{eq} n_j^{eq}+n_i^{eq}\delta_{ij} \delta_{\mathbf{xx'}}.
      \label{secondmoment}
    \end{equation}
    Let us briefly consider the difference between Binomial and Poisson distributed fluctuations. Since the collisions are local, they cannot require knowledge of the system size. So we have to be able to assume a Poisson distribution here. However a finite system size will introduce a correlation between the densities, leading to a distribution that depends on the system size and is not identical to the Poisson distribution.

    There is a fundamental issue that the lattice Boltzmann densities are real numbers, not integers, so strictly they cannot be Poisson distributed. Also there is no accepted generalization of the Poisson distribution to continuous variables. Simply re-interpreting the factorial using a $\Gamma$-function will alter the moments of the distribution function. We will see below that these difficulties become noticeable when we attempt to use the lattice Boltzmann method for very small densities where the discreteness becomes apparent.

    These difficulties arise generally when one attempts to add fluctuations to a continuous theory. Einstein's famous 1905 paper \cite{Einstein} was celebrated for explaining Brownian motion as arising from collisions with discrete solvent molecules, thereby unambiguously showing that even apparently continuous liquids are made up of discrete atoms. Here we attempt to derive a continuous theory that mimics the discrete fluctuating dynamics on length-scales that allow for a continuous treatment.

    To transfer our simple results for the discrete Poisson distribution to our lattice Boltzmann method we will simply require that the first two moments of the discrete and continuous distributions agree. So we demand that
    \begin{align}
     <f_i> &= f_i^{eq},\\
     <f_i f_j> &= f_i^{eq}f_j^{eq}+f_j^{eq}\delta_{ij}.
     \label{ffluct}
    \end{align}
    The basic idea of the following derivation is now to use our knowledge of the distribution of the $<f_i>$ to predict the required fluctuation terms $\xi_i$ that will generate them.

    As a brief aside we want to mention here that this raises an issue of the interpretation of the continuous densities $f_i$. They were originally derived as ensemble averages from discrete lattice gas occupation numbers \cite{higuera}. As such they should not experience fluctuations. Introducing fluctuations now generates a new quantity that does fluctuate as if there were discrete particles but where the occupation numbers remain continuous. This odd construct has the advantage of conserving the freedom of using a much simpler BGK and more flexible collision operator of eqn (\ref{collision}).

    Let us consider the connection between the local equilibrium distribution $f_i^0$  and the global equilibrium distribution $f_i^{eq}$. We can average the lattice Boltzmann eqn. (\ref{LB}) to get
    \begin{align}
    &  <f_i(x+v_i,t+1)> = <f_i(x,t)>+ \nonumber\\&\sum_j \Lambda_{ij}(<f_j^0(x,t)>-<f_j(x,t)>)+<\xi_i(x,t)>,\\
    &  f_i^{eq} = f_i^{eq} +\sum_j\Lambda_{ij}(<f_j^0(x,t)>-f_j^{eq})+0
    \end{align}
    where we used the fact that the expectation values are translationally invariant and that $<\xi_i>=0$, since the noise term is designed to only contain deviations from the mean behavior. We further demand that the collision matrix $\Lambda_{ij}$ be invertible to get
    \begin{equation}
      <f_i^0(x,t)>=f_i^{eq}
      \label{f0feq}
    \end{equation}
    so the expectation values of the local equilibria are the same as the expectation values of the densities. We will use this insight below to derive the global equilibrium distribution from the imposed local equilibrium distribution of the lattice Boltzmann method.

    Next we calculate the expectation value of a lattice Boltzmann equation multiplied with itself, which will give us the noise correlators $<\xi_i \xi_j>$. We get
    \begin{align*}
    &<f_i(x+v_i,t+1)f_j(x+v_j,t+1)>\\
      =&<f_i(x,t)f_j(x,t)>+<f_i(x,t)\Omega_j>
      \nonumber\\&+<\Omega_i f_j(x,t)>+<\Omega_i\Omega_j>.
    \end{align*}
    The first two terms cancel again, and we drop the spatial and temporal dependence for ease of notation. We then get, substituting the BGK collision operator (\ref{collision}),
    \begin{align}
    0 =& \sum_k \Lambda_{jk}<f_i(f_k^0-f_k)>+\sum_l \Lambda_{il}<(f_l^0-f_l)f_j>\nonumber\\
    &+\sum_{kl}\Lambda_{ik}\Lambda_{jl}<(f_k^0-f_k)(f_l^0-f_l)>+<\xi_i\xi_j>.
    \label{corr1}
    \end{align}
    We know (or more exactly require) the moment $<f_i f_j>$ from eqn. (\ref{ffluct}). All that remains is to find $<f_i f_j^0>$ and $<f_i^0 f_j^0>$.

    Up to this point we have not specified what the local equilibrium distribution is. The choice of equilibrium distribution -- and of the conserved quantities -- determines the partial differential equations simulated in the hydrodynamic limit. 

    In most cases the equilibrium distribution is simply a function of the locally conserved variables, i.e. those quantities that cannot be relaxed by the collisions and are therefore the slow variables. The evolution of these slow variables is then discovered in the hydrodynamic limit. 

    We can define the conserved quantities as linear combinations of the $f_i$
    \begin{equation}
      M^c = \sum_i m_i^c f_i
      \label{Mc}
    \end{equation}
    and we then have
    \begin{equation}
      f_i^0(\{M^c\}),
    \end{equation}
    i.e. the local equilibrium is only a function of these conserved quantities. And therefore we know the local equilibrium distribution in terms of the $f_i$.

    Up to this point our considerations have been entirely general. To progress from here we now need to define the local equilibrium distribution function. Here we have only one locally conserved quantity, the density, defined as
    \begin{equation}
      \rho(\mathbf{x},t) = \sum_i f_i(\mathbf{x},t).
    \end{equation}
    To simulate the diffusion equation we demand the following moments for the equilibrium distribution:
    \begin{align}
    \sum_i f_i^0 &= \rho,
      \label{f0_0}\\
    \sum_i f_i^0 v_{i\alpha}&= 0,\\
    \sum_i f_i^0 v_{i\alpha} v_{i\beta}&= \rho \theta \delta_{\alpha\beta}.
      \label{f0_2}
    \end{align}
    In section \ref{hydro} we show that this choice of local equilibrium distribution leads to the diffusion equation in the hydrodynamic limit. Here we are interested in a two-dimensional model, but the derivations are near identical for a one or three dimensional model. The simplest two dimensional velocity set consistent with these moments is
    \begin{equation}
      \{v_i\} = \left\{
      \left(\begin{array}{r}0\\0\end{array}\right),
      \left(\begin{array}{r}1\\0\end{array}\right),
      \left(\begin{array}{r}-1\\0\end{array}\right),
      \left(\begin{array}{r}0\\1\end{array}\right),
      \left(\begin{array}{r}0\\-1\end{array}\right)
      \right\}
    \end{equation}
    which is often also referred to as a D2Q5 model. This leads to the equilibrium distribution
    \begin{equation}
      f_i^0 = \rho w_i
      \label{f0def}
    \end{equation}
    with weights $w_i$ given by
    \begin{equation}
      w_0 = (1-2\theta),\;\;\;\;\;w_{1\ldots4}=\frac{\theta}{2}.
    \end{equation}

    We can now calculate the global equilibrium distribution using eqn. (\ref{f0feq}):
    \begin{align*}
    f_i^{eq} &= <f_i^0>\\
    &= \left<w_i \sum_j f_j\right>\\
    &=  w_i \sum_j f_j^{eq}
    \end{align*}
    Let us define
    \begin{equation}
    \rho^{eq} = \sum_i f_i^{eq}
    \end{equation}
    and we get
    \begin{equation}
      f_i^{eq} = \rho^{eq}w_i.
      \label{feq}
    \end{equation}
    Here the global equilibrium function has exactly the same form as the local equilibrium function. This result does not hold generally, in particular for instance for hydrodynamics models that conserve momentum \cite{kaehlerUnpub}.

    Now we can calculate the equal time correlators we needed for eqn. (\ref{corr1}).
    We obtain
    \begin{align}
    <f_i f_j^0> &=  \left<f_i w_j \sum_k f_k\right>\nonumber\\
    &=\sum_k (f_i^{eq}f_k^{eq}+f_i^{eq}\delta_{ik}) w_j\nonumber\\
    &=f_i^{eq}\rho^{eq}w_j+f_i^{eq}w_j\nonumber\\
    &=(\rho^{eq}+1)\rho^{eq} w_i w_j
    \label{f0f}
    \end{align}
    and
    \begin{align}
    <f_i^0 f_j^0> &= \left<w_i\sum_k f_k w_j \sum_l f_l\right>\nonumber\\
    &=\sum_{kl} w_i w_j (f_k^{eq} f_l^{eq}+f_k^{eq}\delta_{kl})\nonumber\\
    &=(\rho^{eq}+1)\rho^{eq}w_i w_j.
    \label{f0f0}
    \end{align}
    For this simple model both correlators are the same and we have
    \begin{equation}
    <f_i^0 f_j>-<f_if_j>=\rho^{eq}(w_iw_j-w_i\delta_{ij}).
    \end{equation}
    As a side note we would like to point out that earlier derivations of fluctuating lattice Boltzmann methods \cite{adhikari,duenweg,kaehlerPRE} did not know how to treat these correlators and made an assumption equivalent to neglecting the $\delta_{ij}$ term. This resulted in formal results that did not require the fluctuations on conserved moments to vanish. This requirement was then added by hand on physical grounds, an argument which this new derivation avoids, as will become clear below. 

    Using this result in (\ref{corr1}) we obtain
    \begin{align*}
    <\xi_i\xi_j> = \rho^{eq}&\left[\sum_{kl}\Lambda_{ik}\Lambda_{jl}(w_k w_l-w_k\delta_{kl})\right.\\
      &\;\;-\sum_k \Lambda_{jk} (w_iw_k-w_i\delta_{ik})\\
      &\;\;\left. -\sum_l\Lambda_{il} (w_lw_j-w_l\delta_{lj})\right].
    \end{align*}
    This fluctuation dissipation relation is, in some sense, our answer, but it is not a nice answer. We have a full correlation matrix for the noise terms, suggesting that the noise terms are not independent. 

    This is not a surprise, but rather a consequence of the local conservation laws which require
    \begin{equation}
    \sum_i \xi_i = 0
    \end{equation}
    since the fluctuations can't create or destroy mass. 

    This suggests that moving to a different representation in velocity space which separates out the conserved moment may be useful. This moment representation is analogous to what has been used for deriving fluctuating hydrodynamics in the past since the paper by Adhikari \cite{adhikari}. Here we employ the moment transformation rather later in the derivation, which avoids having to perform a rather cumbersome Fourier transform which complicated earlier derivations.

    We now define moments through a general linear transformation
    \begin{equation}
      M^a = \sum_i m_i^a f_i,
    \end{equation}
    which will include the conserved moments mentioned in eqn. (\ref{Mc}), together with a back-transform
    \begin{equation}
      f_i = \sum_a n_i^a M^a
    \end{equation}
    so that we have the relations
    \begin{equation}
      \sum_i n_i^a m_i^b=\delta^{ab}\;\;\;\;\;\sum_a m_i^a n_j^a=\delta_{ij}
    \end{equation}
    and we can move freely between $f$ and $M$ space. We already know that we want $M^0=\rho=\sum_i 1 f_i$, which is equivalent to 
    \begin{equation}
      m_i^0 = (1,1,1,1,1).
    \end{equation}

    One additional consideration for choosing the moments is the collision matrix. If we use a single-relaxation time collision matrix it is always diagonal. In $f$-space it is given by
    \begin{equation}
      \Lambda_{ij}=\frac{1}{\tau}\delta_{ij}
    \end{equation}
    and in moment space it is given by
    \begin{equation}
      \Lambda^{ab} = \sum_{ij} m_i^a \Lambda_{ij} n_j^b = \frac{1}{\tau} \sum_{ij} m_i^a \delta_{ij} n_j^b = \frac{1}{\tau} \delta^{ab}.
      \label{coll}
    \end{equation}
    For a more general multi-relaxation time method we simply demand that $\Lambda^{ab}$ be diagonal:
    \begin{equation}
      \Lambda^{ab} = \frac{1}{\tau^a} \delta^{ab}.
    \end{equation}

    In moment space we then get
    \begin{align}
      <\xi^a \xi^b> = \rho^{eq}&\left[\sum_{ijkl}m_i^a m_j^b\Lambda_{ik}\Lambda_{jl}(w_k w_l-w_k\delta_{kl})\right.\nonumber\\
        &\;\;-\sum_{ijk}m_i^a m_j^b \Lambda_{jk}(w_i w_k-w_i\delta_{ik})\nonumber\\
        &\left.\;\;-\sum_{ijl} m_i^a m_j^b \Lambda_{il} (w_l w_j - w_l\delta_{lj})\right].
      \label{xaxb}
    \end{align}
    Careful inspection reveals that this simplifies if we choose
    \begin{equation}
      n_i^a = m_i^a w_i.
    \end{equation}
    The matrix of moments $m_i^a$ is generated by selecting physically relevant vectors $1$, $v_{ix}$, $v_{iy}$, $v_{ix} v_{ix}-v_{iy}v_{iy}$, and $v_{ix}v_{ix}+v_{iy}v_{iy}$ and employing the Gram-Schmidt orthonormalization procedure with respect to the scalar product
    \begin{equation}
      <m^a m^b>_M = \sum_i m_i^a w_i m_i^b \stackrel{!}{=} \delta^{ab}.
    \end{equation}
    The resulting moment matrix is
    \begin{equation}
        m_i^a = \left(
      \begin{array}{rrrrr}
        1&1&1&1&1\\
        0&\sqrt{\frac{1}{\theta}}&-\sqrt{\frac{1}{\theta}}&0&0\\
        0&0&0&\sqrt{\frac{1}{\theta}}&-\sqrt{\frac{1}{\theta}}\\
        0&\sqrt{\frac{1}{2\theta}} &\sqrt{\frac{1}{2\theta}} &-\sqrt{\frac{1}{2\theta}}   &-\sqrt{\frac{1}{2\theta}}\\
        -\sqrt{\frac{2\theta}{1-2\theta}} &\sqrt{\frac{1-2\theta}{2\theta}} &\sqrt{\frac{1-2\theta}{2\theta}} &\sqrt{\frac{1-2\theta}{2\theta}} &\sqrt{\frac{1-2\theta}{2\theta}}
      \end{array}
      \right)
    \end{equation}
    It may be noted that this matrix differs from the one presented by Li et al. \cite{LiPRE2014} or Le et al. \cite{LePRE2015}. The eigenvectors of the collision matrix in these approaches differ from the current ones only by a factor and additions of conserved eigenvectors. Kaehler et al. \cite{KaehlerTaylor} showed that this ensures that the collision terms are equivalent.

For the equilibrium distribution in moment space we obtain
\begin{equation}
  M^{a,0} = \sum_i m_i^a f_i^{0}.
\end{equation}
For system with mass conservation the first moment is always given by $m_i^0=(1,1,1,1,1)$, and it can be freely inserted as a factor. This implies
\begin{equation}
  M^{a,0} =\sum_i m_i^a f_i^{0} m_i^0=\rho \sum_i m_i^a w_i m_i^0 = \rho \delta^{a0}
\end{equation}
i.e. the local equilibrium density is $\rho$ for the first moment and vanishes for all other moments by virtue of the orthogonalization condition of the moments.

Note that this also implies
\begin{equation}
  \sum_i m_i^a w_i = \delta^{a0}.
\end{equation}

We now obtain for the last term of the noise amplitudes of eqn. (\ref{xaxb})
\begin{align*}
  &\sum_{jk} m_j^b\Lambda_{jk} w_k m_k^0\sum_i m_i^a w_i-\sum_{jk}m_j^b \Lambda_{ik}w_k m_k^a\\
  =& \Lambda^{b0} \rho^{eq} - \Lambda^{ab}\\
  =& \rho (\delta^{a0}\delta^{b0}-1)\delta^{ab} (1/\tau^a)
\end{align*}
and for the first term we get
\begin{align*}
  &\sum_{ik} m_i^a\Lambda_{ik}w_k m_k^0 \sum_{jl}m_j^b\Lambda_{jl}w_l m_l^0\\
  &-\sum_{ijkl} m_i^a\Lambda_{ik} w_k \sum_c m_k^c w_l m_l^c \Lambda_{jl} m_j^b\\
  =&\Lambda^{a0}\Lambda^{b0} - \sum_c \Lambda^{ac}\Lambda^{bc}\\
  =&(\delta^{a0}\delta^{b0}-1) \delta^{ab} (1/(\tau^a)^2).
\end{align*}
With this we get
\begin{equation}
  <\xi^a \xi^b> = \rho^{eq}\frac{2 \tau^a-1}{(\tau^a)^2} (1-\delta^{a0}\delta^{b0})\delta^{ab}.
  \label{xiglobal}
\end{equation}
In this representation the noise terms are diagonal, i.e. no longer correlated, and we explicitly see that the noise amplitude for the conserved mode is zero.

However, this does not quite restrict the noise amplitudes. This is fortunate since 
it seems rather unsatisfactory that the noise should depend on some global average density, a bit of information that should not be available to a local collision operator. This was first argued by Kaehler et al. \cite{kaehlerPRE}.Let us define $<\cdots>|_\rho$ as the ensemble average over a all states for a cell with local density $\rho$. We can then write a local noise term of the form
\begin{equation}
  <\xi^a \xi^b>|_\rho = \rho \frac{2 \tau^a-1}{(\tau^a)^2} (1-\delta^{a0}\delta^{b0})\delta^{ab}.
  \label{xilocal}
\end{equation}
While local noise had been proposed before by Kaehler, we are now able to show that local noise is consistent with (\ref{xiglobal}), which was not possible in earlier derivations. 
\begin{align*}
  <<\xi^a \xi^b>|_\rho> &= <\rho> \frac{2 \tau^a-1}{(\tau^a)^2} (1-\delta^{a0}\delta^{b0})\delta^{ab}\\
  &= \rho^{eq} \frac{2 \tau^a-1}{(\tau^a)^2} (1-\delta^{a0}\delta^{b0})\delta^{ab}
\end{align*}
which is now consistent with (\ref{xiglobal}),  and this local formulation does not require knowledge of the whole system to determine the local noise amplitude. We will show below that local noise is required to fully recover the correct behavior of the system, particularly for non-equilibrium systems with large variations in the density.

\section{\label{hydro} Hydrodynamic limit}
We claimed earlier that the choice of equilibrium distribution given by the moments (\ref{f0_0}--\ref{f0_2}) will lead the lattice Boltzmann method to simulate a diffusion equation. In this section we will now derive the hydrodynamic limit of the fluctuating lattice Boltzmann method and show that we recover a fluctuating diffusion equation.

This derivation follows the approach developed by Kaehler et al. \cite{KaehlerTaylor}. We write the lattice Boltzmann eqn. (\ref{LB}) in terms of a Taylor expansion to get
\begin{equation}
  (\partial_t + v_{i\alpha}\partial_\alpha) f_i + \frac{1}{2} (\partial_t + v_{i\alpha}\partial_\alpha)^2 f_i = \sum_j \Lambda_{ij} (f_j^0-f_j)+\xi_i.
\label{LBt}
\end{equation}
We can use this equation to express the distribution $f_i$ in terms of the local equilibrium distribution and derivatives. Using that the collision matrix is invertible we get to first order
\begin{equation}
  f_i = f_i^0 - \sum_j \Lambda^{-1}_{ij} (\partial_t + v_{j\alpha}\partial_\alpha) f_j+\sum_j\Lambda^{-1}_{ij} \xi_j.
\label{ff0}
\end{equation}
Now we reinsert this expression into our expanded lattice Boltzmann eqn. (\ref{LBt}) to get
\begin{align}
 & (\partial_t+v_{i\alpha}\partial_\alpha)\left(f_i^0+\sum_j \Lambda^{-1}_{ij}\xi_j\right)\nonumber\\&
  - (\partial_t+v_{i\alpha}\partial_\alpha)\sum_j \left(\Lambda^{-1}_{ij}-\frac{1}{2}\delta_{ij}\right) (\partial_t + v_{j\beta}\partial_\beta) f_j\nonumber\\
  =& \sum_j \Lambda_{ij}(f_j^0-f_j)+\xi_i.
\end{align}
We then reinsert eqn. (\ref{ff0}) to also replace the remaining occurrence of $f_i$ on the left hand side and
sum the resulting equation over $i$ to obtain the hydrodynamic equation for the evolution of the density $\rho$. We get
\begin{align}
 & \partial_t \rho +\partial_\alpha\sum_{ij}v_{i\alpha} \Lambda^{-1}_{ij}\xi_j
  \nonumber\\&
  - \sum_i(\partial_t+v_{i\alpha}\partial_\alpha)\sum_j \left(\Lambda^{-1}_{ij}-\frac{1}{2}\delta_{ij}\right)
  \nonumber\\&\;\;
  (\partial_t + v_{j\beta}\partial_\beta) \left(f_j^0+\sum_k\Lambda_{jk}^{-1}\xi_k\right)
  = 0.
  \label{LBexp2}
\end{align}

The collision matrix should be isotropic, i.e. relaxation should be invariant under rotation. This restricts the collision matrix (\ref{coll}) to be of the form
\begin{equation}
\Lambda^{ab} = \mbox{diag}(\tau^\rho,\tau^j,\tau^j,\tau^n,\tau^s).
\end{equation}
Now let us consider $v_{i\alpha} \Lambda_{ij}^{-1}$. Noting that we constructed the moment matrix such that $v_{i\alpha}=\sqrt{\theta}m_i^{\alpha+1}$ we can write
\begin{align}
 \sum_i v_{i\alpha}\Lambda_{ij}^{-1}&=\sum_i \sum_{ab}\sqrt{\theta}m_i^{\alpha+1} n_i^a\tau^a\delta^{ab} m_j^b\nonumber\\
 &=\sqrt{\theta}\sum_b \delta^{\alpha+1,b}\tau^bm_j^b\\
 &=\tau^jv_{j\alpha}.
\end{align}
This shows that the velocity $v_{i\alpha}$ is a left eigenvector of the inverse collision matrix with eigenvalue $\tau^j$. Similarly $1_i$ (i.e. a vector consisting entirely of ones) is also a left eigenvector with the eigenvalue $\tau^\rho$. Note that this eigenvalue is arbitrary since the density is conserved due to eqn. (\ref{f0_0}). This implies that all terms containing this arbitrary factor must cancel from the hydrodynamic equations.

For the second noise term we need to consider
\begin{align}
  \sum_{ij} v_{i\alpha} \Lambda_{ij}^{-1}v_{j\beta}\Lambda_{jk}^{-1}
  &= \frac{1}{\tau^j} \sum_j v_{j\alpha}v_{j\beta}\Lambda_{jk}^{-1}.
\end{align}
For our D2Q5 model $v_{ix}v_{iy}=0$ by construction, so we only need to worry about the $v_{ix}v_{ix}$ and the $v_{iy}v_{iy}$ term in the expression above. We can write these in terms of the left eigenvectors as
\begin{equation}
v_{ix}v_{ix} = \theta \left(m_i^0+\sqrt{\frac{1}{2\theta}}m_i^4+\sqrt{\frac{1-2\theta}{2\theta}}m_i^5\right)
\end{equation}
and we get
\begin{equation}
  \sum_i v_{ix}v_{ix} \Lambda_{ij}^{-1} = \tau^\rho m_i^0+\sqrt{\frac{1}{2\theta}}\tau^n m_i^4+\sqrt{\frac{1-2\theta}{2\theta}}\tau^s m_i^5.
\end{equation}
If we now define the macroscopic noise terms as
\begin{align}
  \eta_\alpha = \tau^j \sum_i v_{i\alpha}& \xi_i=\tau^j \sqrt{\theta} \xi^{1+\alpha}\\
  \chi_{\alpha\beta}=
  \delta_{\alpha\beta}\sum_i&\left[\tau^s\frac{1-2\theta}{2}(v_{ix}v_{ix}+v_{iy}v_{iy})
  \right.\nonumber\\&\left.+\tau^n(-1)^{\delta_{\alpha y}}\frac{1}{2}(v_{ix}v_{ix}-v_{iy}v_{iy})\right]\xi_i.
\end{align}
We now get for (\ref{LBexp2})
\begin{align}
  &\partial_t \rho + \partial_\alpha \eta_\alpha + \partial_t (\tau^\rho-\frac{1}{2}) (\partial_t\rho+\partial_\beta \eta_\beta)
  \nonumber\\&+\partial_\alpha (\tau^j-\frac{1}{2})\left[\partial_t \eta_\alpha+\partial_\beta (\rho \theta\delta_{\alpha\beta}+\chi_{\alpha\beta})\right]=0.
\end{align}
The third term here contains of a time derivative of the first two terms and therefore is of third order. The appearance of the irrelevant relaxation time $\tau^\rho$ implies that this term will vanish not only to third, but to all orders. It is therefore neglected. We then arrive at the macroscopic fluctuating diffusion equation
\begin{equation}
  \partial_t \rho + \partial_\alpha \eta_\alpha
  +\partial_\alpha \frac{D}{\theta}(\partial_\alpha \rho \theta+ (\partial_t \eta_\alpha+\partial_\beta \chi_{\alpha\beta}))=0 
\end{equation}
where we introduced the diffusion constant
\begin{equation}
  D = \left(\tau^j-\frac{1}{2}\right)\theta
\end{equation}
Now formally it may look as if $\partial_\alpha\eta_\alpha$ was of order $O(\partial)$. However, in equilibrium this random part of the diffusion current will be on average equal to the deterministic restoring part of the diffusion current $\partial_\alpha \partial_\alpha (\rho\theta)$ and it will therefore also be of order $O(\partial^2)$. Note that this is equivalent to the way one shows that the forcing term in a lattice Boltzmann method has to be of order $O(\partial)$, see e.g. Li et al. \cite{LiPRE}. 
Therefore the terms
\begin{equation}
\partial_t \eta_\alpha + \partial_\beta \chi_{\alpha\beta} = O(\partial^3)
\end{equation}
are of third order.

We then get
\begin{equation}
  \partial_t \rho + \partial_\alpha \eta_\alpha
  +\partial_\alpha \frac{D}{\theta}\partial_\alpha \rho\theta=0.
  \label{DiffEqn}
\end{equation}
For a constant temperature $\theta$ this leaves us with the standard fluctuating diffusion equation
\begin{equation}
  \partial_t \rho + \partial_\alpha \eta_\alpha
  +\partial_\alpha D\partial_\alpha \rho=0 
\end{equation}
where the noise current $\eta_\alpha$ obeys
\begin{align}
  <\eta_\alpha> &= 0\\
  <\eta_\alpha \eta_\beta> &= \tau^j \tau^j \theta <\xi^{1+\alpha}\xi^{1+\beta}>\nonumber\\
  &= (2\tau^j-1) \rho^{eq} \theta \delta_{\alpha\beta}\nonumber\\
  &= 2 \rho^{eq} D \delta_{\alpha\beta}
\end{align}
for the global noise implementation and 
\begin{align}
  <\eta_\alpha> &= 0\\
  <\eta_\alpha \eta_\beta> &= 2 \rho D \delta_{\alpha\beta}
\end{align}
for the local noise implementation.

\section{Simulation results}
To fully determine the algorithm we need to briefly discuss our implementation of the noise of eqn. (\ref{xiglobal}) and (\ref{xilocal}).
For simplicity we implement the uncorrelated $\xi^a$ noise terms with a flat distribution:
\begin{equation}
P(\xi^a) = \left\{ 
\begin{array}{cl}
\frac{1}{2d^a} & \mbox{for }-d^a<\xi^a<d^a\\
0   & otherwise
\end{array}
\right.
\end{equation}
with
\begin{equation}
d^a = \sqrt{3\rho^{eq}\frac{2\tau^a-1}{(\tau^a)^2}}
\end{equation}
for the global noise implementation and 
\begin{equation}
d^a = \sqrt{3\rho\frac{2\tau^a-1}{(\tau^a)^2}}
\end{equation}
for the local noise implementation. For negative densities, which are unphysical but may appear in this numerical method, this is treated as if the density was zero. This fully determines our algorithm.

Now we need to test the consistency of the proposed algorithm and evaluate its limitations. We have shown that the noise terms of eqn. (\ref{xiglobal}) are necessary to reproduce imposed correlators of the distribution functions in equilibrium given by eqn. (\ref{ffluct}). However, it is not clear that this is sufficient. Past implementations for fluctuating hydrodynamics have shown noticeable deviations from this expectation \cite{kaehlerPRE,kaehlerThesis}.

\begin{figure}
  \includegraphics[width=\columnwidth,clip=true]{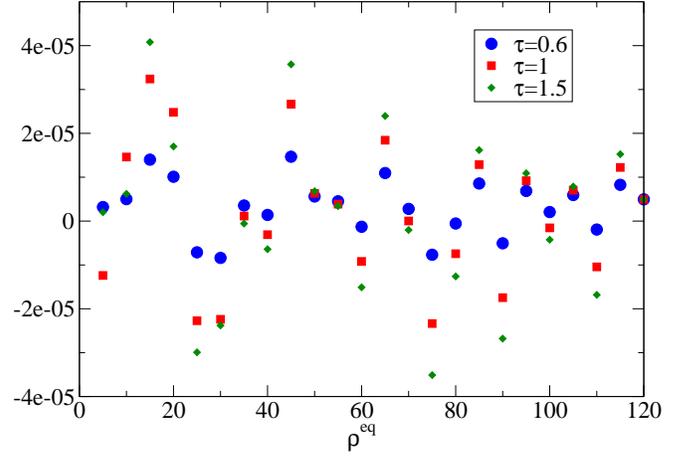}
  \caption{Difference of the predicted equilibrium distribution of eqn. (\ref{f0def}) and the measured equilibrium distribution as a function of the average density for different relaxation times. The results are for $\theta=1/3$ averaged over $10^6$ iterations and show excellent agreement. Results shown are for the local noise implementation, the global noise implementation shows similarly small errors.}
  \label{Figf0}.
\end{figure}

Firstly we need to check that the equilibrium distribution predicted by eqn. (\ref{feq}) is indeed reproduced by our method. We compare this with a time and space average
\begin{equation}
  <f_i>_{exp} = \frac{1}{T}\sum_{t=1}^T \frac{1}{L^2}\sum_\mathbf{x} f_i(\mathbf{x},t)
\end{equation}
with the equilibrium distribution for different average densities in Fig. \ref{Figf0}. Here we sum over the whole two dimensional lattice, which for simplicity of notation is supposed to be square and have $L$ points in each of the two directions. We observe excellent agreement independent of the relaxation times and even for low densities. 

A small note: for hydrodynamic lattice Boltzmann methods there is a difference between the effective temperature for local and global equilibrium distributions. However, this effect is closely linked to momentum conservation, which is absent in the model considered here.

Next we examine the second order moments of the distribution functions
\begin{equation}
  d_{ij}= \frac{<f_i f_j>-f_i^{eq}f_j^{eq}}{\sqrt{f_i^{eq}f_j^{eq}}} \stackrel{?}{=}\delta_{ij}
  \label{dij}
\end{equation}
where we re-expressed eqn. (\ref{ffluct}) to give an expression that is intuitive to test.
We numerically evaluate $d_{ij}$ by averaging over the whole lattice and over a large number of iterations as
\begin{equation}
  d_{ij}^{exp} = \frac{1}{T} \frac{1}{L^2}\sum_{t=1}^T \sum_\mathbf{x} \frac{f_i(\mathbf{x},t)f_j(\mathbf{x},t)-\rho^{eq} w_i \rho^{eq} w_j}{\rho^{eq}\sqrt{w_i w_j}}.
  \label{dijexp}
\end{equation}

\begin{table}
  \subfloat[3x3 lattice points\label{correllator3x3}]{
    \begin{minipage}{\columnwidth}
    $$
    \begin{array}{rrrrr}
      0.962681 & -0.026427 & -0.026222 & -0.026236 & -0.026376\\
      -0.026427 & 0.981131 & -0.018669 & -0.018514 & -0.018678\\
      -0.026222 & -0.018669 & 0.981129 & -0.017702 & -0.018877\\
      -0.026236 & -0.018514 & -0.017702 & 0.981634 & -0.018937\\
      -0.026376 & -0.018678 & -0.018877 & -0.018937 & 0.980849
    \end{array}
    $$
    \end{minipage}
  }\newline
  \subfloat[10x10 lattice points\label{correllator10x10}]{
    \begin{minipage}{\columnwidth}
    $$
    \begin{array}{rrrrr}
      0.996554 & -0.002441 & -0.002380 & -0.002300 & -0.002350\\
      -0.002441 & 0.998277 & -0.001719 & -0.001644 & -0.001642\\
      -0.002380 & -0.001719 & 0.998349 & -0.001680 & -0.001739\\
      -0.002300 & -0.001644 & -0.001680 & 0.998543 & -0.001705\\
      -0.002350 & -0.001642 & -0.001739 & -0.001705 & 0.998333
    \end{array}
    $$
    \end{minipage}
  }\newline
  \subfloat[100x100 lattice points\label{correllator100x100}]{
    \begin{minipage}{\columnwidth}
    $$
    \begin{array}{rrrrr}
      0.999968 & -0.000024& -0.000035& -0.000024& -0.000013\\
      -0.000024& 0.999978& -0.000017& -0.000014& -0.000016\\
      -0.000035& -0.000017& 0.999996& -0.000021& -0.000012\\
      -0.000024& -0.000014& -0.000021& 0.999963& -0.000023\\
      -0.000013& -0.000016& -0.000012& -0.000023& 0.999988
    \end{array}
    $$
    \end{minipage}
  }
  \caption{Estimation for $d_{ij}$ from eqn. (\ref{dijexp}) for different lattice sizes over $1.7\cdot 10^{6}$ iterations. Other simulation parameters were kept constant at $\tau^j=\tau^s=\tau^n=1$, $\theta=1/3$, $\rho^{eq}=120$ and the results were averaged over $1.7\cdot 10^{6}$ iterations. The noise amplitude was given by local noise of eqn. (\ref{xilocal}). Global noise results are similar.}
  \label{correlator}
\end{table}

Results of this averaging are shown in Table \ref{correlator}. 
We observe that the second equality in eqn. (\ref{dij}) is not exactly fulfilled. However, the agreement gets better the larger the lattice is. In eqn. (\ref{Poisson}) we assumed that the $n_i$ are Poisson distributed, which is only true when we are dealing with an infinite system. For a finite system we would have a binomial distribution. Physically the origin of this difference is a slight correlation of the distribution function on finite lattices because of global conservation laws, similar to what Ollila et al. \cite{Ollila} observed for momentum conservation. In particular the total density is conserved
\begin{equation}
  \sum_x \sum_i f_i(\mathbf{x},t) = L^2 \rho^{eq}.
\end{equation}
which implies that the $f_i$ are correlated.
The importance of this correlation then depends on the finite system size, and will vanish in the limit of large systems.

\begin{figure}
  \includegraphics[width=\columnwidth,clip=true]{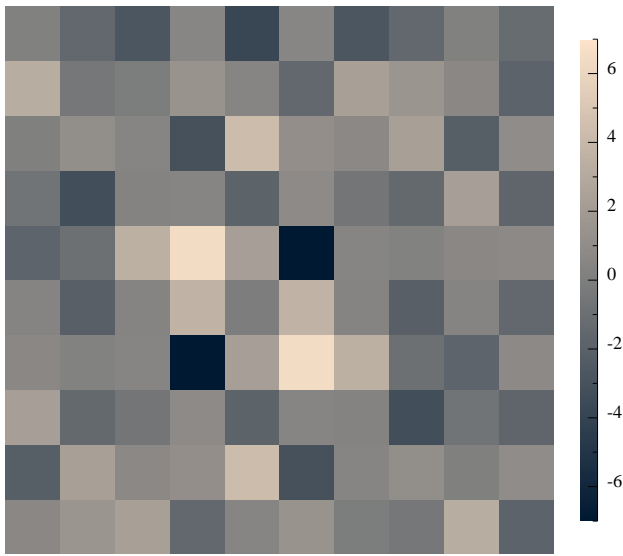}
  \caption{Deviation of Fourier modes from eqn. (\ref{FourierTest}) for a lattice of $L=10$ averaged for $T=336\times 10^{6}$ shows no discernable structure. Note that each of the $f(\mathbf{k},t)f(-\mathbf{k},t)$ to be averaged has values varying between about $\pm 50,000$, so the remaining scale of $\pm 7$ after averaging is very small.}
  \label{FourierFig}
\end{figure}

A useful way to separate out the globally conserved modes from non-conserved ones is to express the densities in terms of their Fourier modes.  
We define the discrete Fourier sums for the densities as
\begin{equation}
  f_i(\mathbf{k},t) = \sum_x \exp\left(\frac{2\pi}{L}\mathbf{k.x}\right) f_i(\mathbf{x},t)
\end{equation}
where $L$ is the total number of lattice points in each spatial direction (assuming a square lattice for ease of notation). We then find a structure factor for the densities~$f_i$:
\begin{align}
S_{ij}(\mathbf{k})  =&<f_i(\mathbf{k},t)f_j(-\mathbf{k},t)>\nonumber\\
  =& \sum_\mathbf{x} \sum_{\mathbf{x'}}\exp\left(\frac{2\pi}{L}\mathbf{k}.(\mathbf{x}-\mathbf{x}')\right) <f_i(\mathbf{x},t)f_j(\mathbf{x}',t)>\nonumber\\
  =&\sum_\mathbf{x} \sum_{\mathbf{x}'}\exp\left(\frac{2\pi}{L}\mathbf{k}.(\mathbf{x}-\mathbf{x}')\right)
  [f_i^{eq}f_j^{eq}+f_i^{eq} \delta_{ij} \delta_{\mathbf{xx}'}]\nonumber\\
  =&\delta_{\mathbf{k}0} L^4 f_i^{eq}f_j^{eq}+L^2 f_i^{eq} \delta_{ij}
\label{fkfk}
\end{align}
which should be flat, except at $k=0$. This result is also not expected to be entirely correct, since we have mass conservation in our finite system. In contrast to the results for the $d_{ij}$, however, the problem caused by mass conservation is limited to the $k=0$ mode. We know that mass conservation requires
\begin{align}
  &  \sum_{ij} f_i(\mathbf{k}=0,t) f_j(\mathbf{k}=0,t)\nonumber\\
  =& \sum_{ij}\sum_x f_i(\mathbf{x},t) \sum_\mathbf{x'} f_j(\mathbf{x}',t)\nonumber\\
  =& L^4 (\rho^{eq})^2
\end{align}
whereas eqn. (\ref{fkfk}) predicts $L^4(\rho^{eq})^2+L^2\rho^{eq}$. We therefore predict the full structure factor for the $f_i$ to be
\begin{align}
  S_{ij}^{th}(\mathbf{k}) = \delta_{\mathbf{k}0}L^4 f_i^{eq} f_j^{eq}+ (1-\delta_{\mathbf{k}0}) L^2 f_i^{eq}\delta_{ij}.
\end{align}

In Fig. \ref{FourierFig} we show the deviation
\begin{align}
  S_{ij}^{diff}(\mathbf{k})= \frac{1}{T}\sum_{t=0}^{T} f_i(\mathbf{k},t) f_j(-\mathbf{k},t)-S_{ij}^{th}(\mathbf{k})
  \label{FourierTest}
\end{align}
and see that, unlike in previous fluctuating methods, we were unable to detect any remaining spurious structure. 

The results so far show that we are able to recover the expected averages and variances for the densities $f_i$, at least in the limit of infinite system size or in Fourier representation for non-zero Fourier modes. These are recovered for noise amplitudes that are dependent on a local density of eqn. (\ref{xilocal}) or on the global density of eqn. (\ref{xiglobal}), and the differences between these two implementations are insignificant.

\subsection{Limit of low density}
As previously mentioned one might expect this method to fail in the limit of low density. When lattice Boltzmann densities sum up to only a few particles or even a fraction of a particle the corresponding Poisson distribution no longer resembles a continuous distribution. However, it is important to know the limits of a numerical method. We therefore resolved to determine the density at which our method ceases to give reasonable results. 

\begin{figure}
  
  \subfloat[Global Noise]{\includegraphics[width=\columnwidth,clip=true]{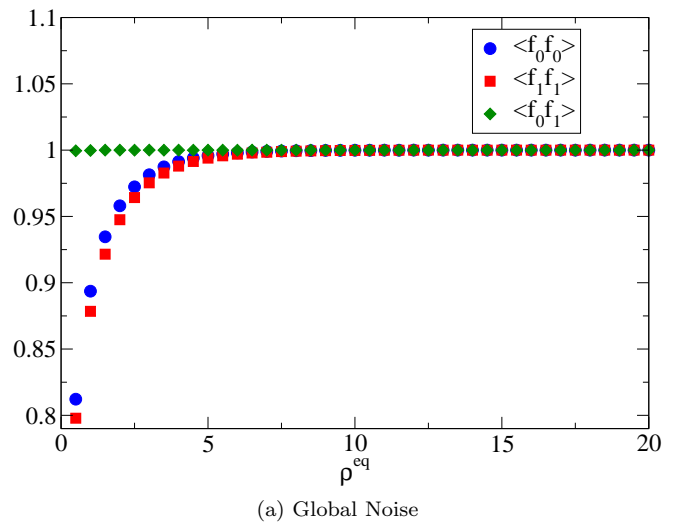}}

  \subfloat[Local Noise]{\includegraphics[width=\columnwidth,clip=true]{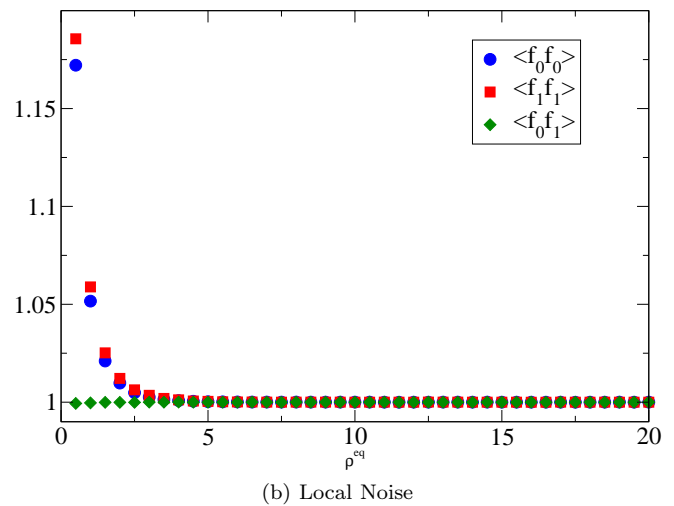}}

  \caption{Correlators $<f_i f_j>$ divided by the predicted result of eqn. (\ref{ffluct}) as a function of the mean densities. Ideally this expression would have a constant value of 1. Unsurprisingly for low densities some deviations are observed. It is noted that the deviation for the global noise amplitude sets in below $\rho^{av}\approx 7$, whereas it only sets in at about $\rho^{eq}\approx 3$ for the local noise implementation, and the agreement is quantitatively better.}
  \label{LowDensity}
\end{figure}

We examined the correlators $<f_i(x,t) f_j(x,t)>$ for increasingly lower densities. The result is shown in Fig. \ref{LowDensity}. Even for low average densities the second moments for the density are recovered with surprising fidelity down to very small average numbers of particles. The error remains below $1\%$ for an average density of $\rho^{eq}$ of about 7 for the global noise implementation of eqn. (\ref{xiglobal}), but is noticeably better for the local noise implementation of eqn. (\ref{xilocal}). Here the threshold of $1\%$ deviation is reached only at a density of $\rho^{eq}\approx 3$. At this density about $5\%$ of sites will have negative densities, where we suspended the fluctuations. It is also interesting to note that global noise leads to smaller fluctuations, whereas local noise increases the fluctuations at low densities. We did not investigate the exact reasons for these deviations since they are small, and approach a $10\%$ error only for densities of less than one particle per lattice site, where one should no longer expect the continuous description to give good description of fluctuations.

\begin{figure}
   \subfloat[$\rho^{eq}=30$]{\includegraphics[width=\columnwidth,clip=true]{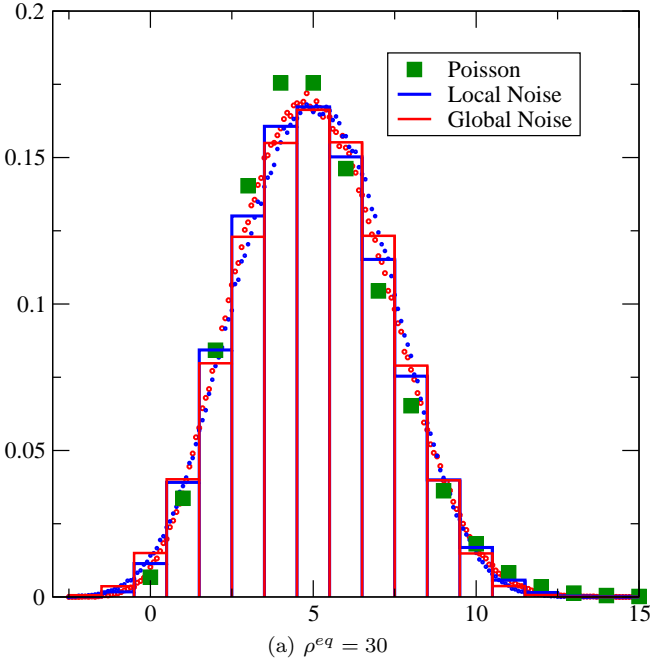}}

\subfloat[$\rho^{eq}=1$ \label{LDD2}]{\includegraphics[width=\columnwidth,clip=true]{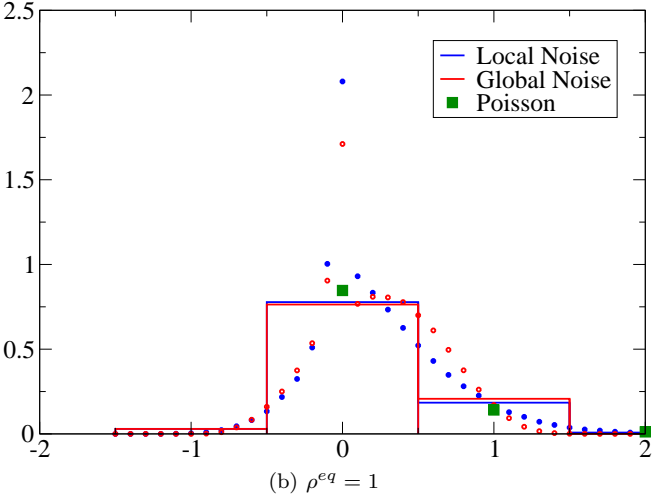}}

   \caption{Comparison of observed distribution functions for local and global noise amplitudes to the Poisson distribution. We show the distribution for $f_1$ for (a) $f_1^{eq}=5$ (for $\rho^{eq}=30$) and (b) $f_1^{eq}=1/6$ (for $\rho^{eq}=1$). }
\label{LowDensityDist}
\end{figure}

\begin{table}
  
  \subfloat[$\rho^{eq}=30$ \label{tab2_1}]{\begin{tabular}{r|rrr}
    &$<f_1>$ &$<f_1^2>$&$<f_1^3>$\\
    \hline
    Poisson & 5.000&30.000&205\\
    Global noise & 4.997 &29.971 &200\\
    Local noise & 5.001 &30.005 &202
  \end{tabular}}
  
  \subfloat[$\rho^{eq}=1$ \label{tab2_2}]{\begin{tabular}{r|rrr}
    &$<f_1>$ &$<f_1^2>$&$<f_1^3>$\\
    \hline
    Poisson & 0.1666&0.1944&0.254\\
    Global noise & 0.1711 & 0.1710 &0.091\\
    Local noise &  0.1666& 0.2071 &0.164
  \end{tabular}}
  
  \caption{The first three moments of the distributions of Fig. \ref{LowDensityDist}.}
  \label{tab2}
\end{table}

We performed a closer comparison of the distribution of the densities. For the case of $f_1$ this analysis is shown in Fig. \ref{LowDensityDist}. We generated histograms with the local and global noise implementations with a resolution of 1 particle and 1/10 of a particle and compared them to the Poisson distribution.

For a total density of $\rho^{eq}=30$ we find fairly good agreement, although the histograms do not exactly match up with each other or the Poisson distribution. Most obviously we see that there is a finite contribution from negative $f_1$. A direct comparison of the first three moments of these distribution functions is shown in Table \ref{tab2_1}. We see that the first two moments match up nicely, with the error being slightly smaller for the local noise implementation, which would have been expected from the results presented in Fig. \ref{LowDensity}.

So far we have only discussed the first two moments since those were the moments that our theory intended to match. However it is reasonable to also look at the third moment, even though there is no theoretical analysis that claims these moments should match. We see in Table \ref{tab2} that the agreement for these third moments is indeed noticeably worse, but the trend of a better fit for the local noise continues. 

Secondly we examined the distribution of $f_1$ for only one particle per lattice site on average. This is shown in Fig. \ref{LDD2}. The Poisson distribution consists essentially only of empty lattice cells and cells with one particle. The continuous generalizations of the Poisson distribution generated by the global and local noise algorithm differ slightly from each other, but both share the curious feature of an apparent singular contribution for a distribution of exactly zero. We do not currently understand this feature. It is not due to the cessation of noise application at zero density, since this only applies to the local noise algorithm, not the global one. The matching of the moments to the Poisson distribution, shown in Table \ref{tab2_2}, is no longer as good. The first moments still match, but there is a 10\% deviation for the second moment in the local noise implementation, and slightly more for the global one. The third moments now vary by much more, and the match between the local noise implementation to the Poisson distribution is much better that the global noise one.

\begin{figure}
  
  \subfloat[Global noise]{\includegraphics[width=\columnwidth,clip=true]{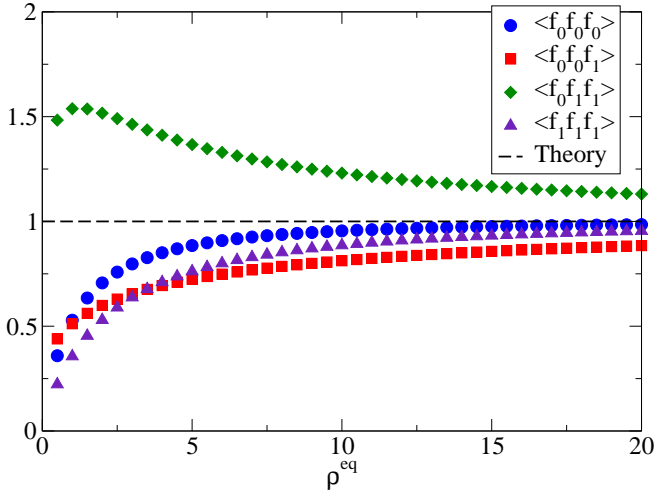}}

  \subfloat[Local noise]{\includegraphics[width=\columnwidth,clip=true]{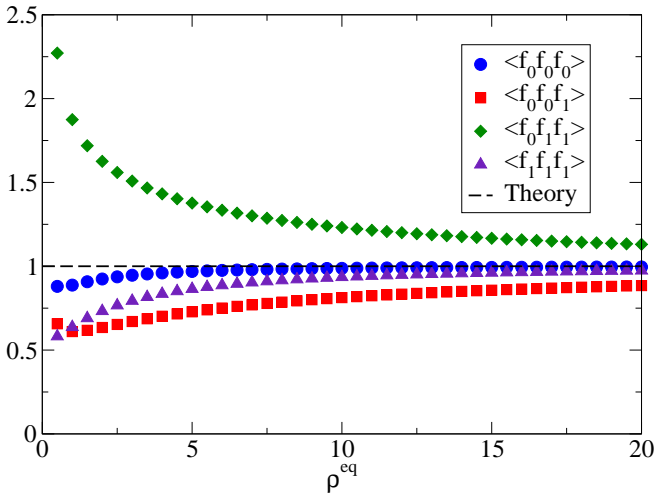}}

  \caption{Third moment correlators divided by the theoretical prediction of eqn.~(\ref{fff}) as a function of the mean densities for both global and local noise. The mixed terms ($<f_0f_0f_1>$, $<f_0f_1f_1>$) have more pronounced deviations from our expected value than the terms examined in Table~\ref{LowDensityDist}.}
  \label{LowDensityThird}
\end{figure}

In the previous analysis we have only looked at one third order moment $<f_1^3>$. For completeness we also examined three other third order moments, i.e. $<f_0^3>$, $<f_0^2 f_1>$ and $<f_0 f_1^2>$.  Eqn.~(\ref{f3}) gives this prediction for these third order moments
\begin{align}
  & < f_i f_j f_k> \nonumber\\
  =& f_i^{eq}f_j^{eq}f_k^{eq}+f_i^{eq}f_j^{eq}\delta_{ik}+f_j^{eq}f_k^{eq}\delta_{ij}+f_i^{eq}f_k^{eq}\delta_{jk}\nonumber\\
  &+f_i^{eq}\delta_{ijk}.
  \label{fff}
\end{align} 
The third moment of the Poisson distribution (a moment that we did not attempt to enforce) shows moderate deviations from the values expected for the Poisson distribution. This is shown in Fig. \ref{LowDensityThird}. For larger densities the agreement increases, as might be expected from the central limit theorem, since the Poisson distribution becomes a Gaussian distribution for large average densities. 

We conclude here that, surprisingly, the method continues to reproduce the expected fluctuations, even for very low densities, and becomes inaccurate only for average densities of about 3 particles per lattice site on average.

\subsection{Non-homogeneous systems}
So far we have only examined systems that are homogeneous and in equilibrium. 
Next we will look at two systems where the dynamics are non-homogeneous. To make the effect as abrupt as possible we examine two domains with a sharp interface where either the temperature or the mobility is different which is highly relevant for the simulation of phase-separation fronts \cite{FoardPRE1,FoardPRE2}.

\begin{figure}
  \subfloat[Two temperatures $\theta_1=1/3$ and $\theta_2=1/6$.\label{ItoStrat1}]{\includegraphics[width=\columnwidth,clip=true]{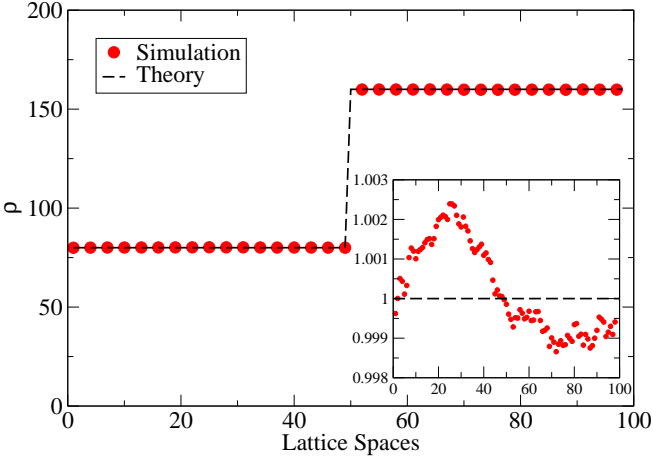}}
  
  \subfloat[Two diffusion constants $D=1/30$ and $D=1/3$.\label{ItoStrat2}]{\includegraphics[width=\columnwidth,clip=true]{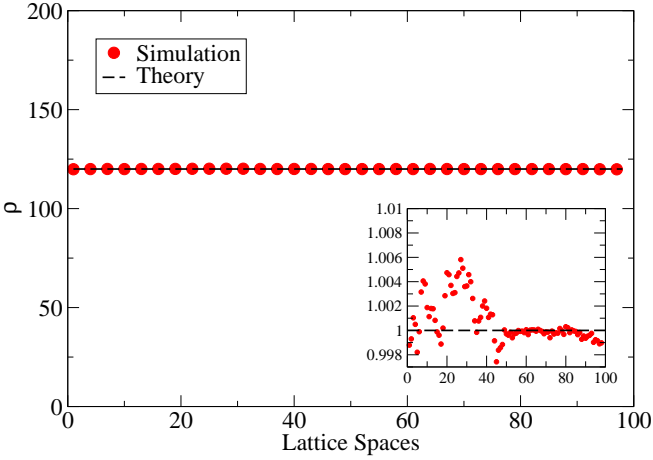}}
  
  \caption{Effect of different mobilities (a) and different temperature (b) in two different regions on the equilibrium behavior of the system. In the case of different temperatures we find a density difference, whereas the case of different mobilities leaves the densities constant. The insets show the numerical solution divided by the analytical solution.}
  \label{ItoStrat}
\end{figure}

First we examined a system where two regions are held at different imposed temperatures $\theta_1$ and $\theta_2$. Here particles will move more quickly in the region of high $\theta$ and therefore spend less time in regions of high temperature. Averaging the hydrodynamic eqn. (\ref{DiffEqn}) we obtain the standard diffusion equation in one dimension
\begin{equation}
  \partial_t \rho = \partial_x \frac{D}{\theta} \partial_x (\rho \theta).
\label{DiffAve}
\end{equation}
Quantitatively we can look at the stationary solution of this equation which implies that the two different average densities in the two regions are given by
\begin{equation}
  <\rho_1 \theta_1> = <\rho_2\theta_2>.
\end{equation}
This situation is shown in Fig. \ref{ItoStrat1}, and we see that the average densities in the two domains are recovered without any noticeable artifacts at the sharp temperature boundary.

The second situation we consider is a system where two regions have different diffusion constants. For this system eqn. (\ref{DiffAve}) predicts a constant density. This situation is related to the so-called Ito-Stratonovich dilemma. A single particle in this situation would undergo a random walk, and the likelihood to find it in either region (of equal size) should be the same. However, since the particle is moving more quickly in the region with high mobility the path length inside this region has to be longer. This is only possible if there is an preferential reflection at the interface between the two regions into the region with the higher mobility. 

The way to modify the diffusion constant in a region is through the current relaxation time $\tau^j$, which now becomes a function of space. The results of Fig. \ref{ItoStrat2} show that the expected behavior is recovered by our lattice Boltzmann method. We should also note that the presence of fluctuations is incidental to this problem. The same behavior persists for the deterministic diffusion equation simulated by lattice Boltzmann, but it worthwhile to note that this property is unaffected by the introduction of noise.

\begin{figure}
\includegraphics[width=\columnwidth,clip=true]{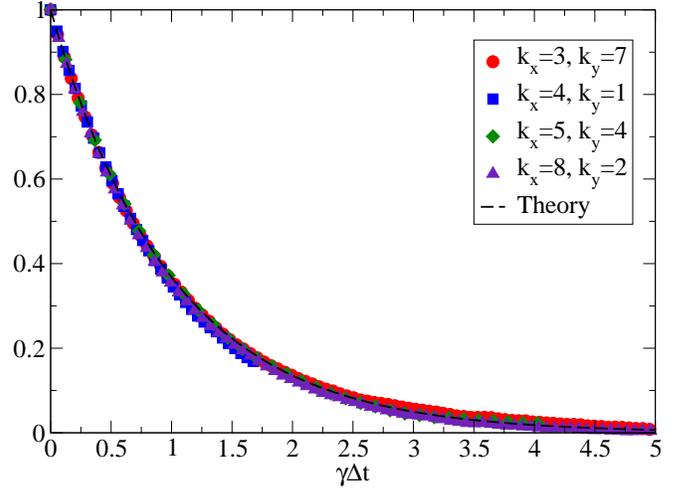}
  \caption{Time correlators of eqn. (\ref{TimeCorrEqn}) for an 81x81 lattice and $\rho^{eq}=120$, $D=1/6$ for different $k$ values.. Values have been averaged over $2\cdot10^6$ iterations.}
\label{TimeCorr}
\end{figure}

\subsection{Equilibrium dynamics}
A simple heuristic argument for the time-correlation function relies on the Fourier transform of the diffusion equation. To predict the behavior of the discrete system we can start by examining the lattice Boltzmann eqn. (\ref{LB}). This representation becomes simple for the special collision operator of $\Lambda_{ij}=1$, i.e. the case where all relaxation times are one. In that case we get
\begin{equation}
  f_i(x+v_i,t+1)=w_i \sum_i f_i(x,t)+\xi_i(x,t)
\end{equation}
We can translate this equation by $-v_i$ and sum it over $i$. We then subtract the local density at time $t$ to obtain
\begin{align}
  &\rho(x,t+1)-\rho(x,t)\nonumber\\
  =& \sum_i w_i (\rho(x-v_i,t)-\rho(x,t))+\sum_i \xi_i(x-v_i,t).
\end{align}
For the D2Q5 model employed in our example we can then write a discrete Fourier transform of this equation to calculate the decay times for a particular Fourier mode (neglecting noise terms for now, following Forster's \cite{Forster1975} phenomenological approach).

We believe that it is possible to perform a fully consistent analysis of time correlations, only relying on the correlation of equal time $<f_i(x,t)f_j(x',t)>=f_i^{eq}f_j^{eq}+f_i^{eq}\delta_{ij}\delta_{xx'}$ by inserting the evolution equation. However, initial investigation of this approach show somewhat lengthy results, that would unreasonably expand this already lengthy paper. 

We change our notation and write out the position vector as $\mathbf{x}=(x,y)$, as the results are not fully isotropic, and therefore cannot be written in vector form. We have
\begin{align}
  &\sum_{x,y} e^{i2\pi(\frac{k_x x}{L_x}+\frac{k_y y}{L_y})} [\rho(x,y,t+1)-\rho(x,t)]\nonumber\\
  =& D \sum_x e^{i2\pi(\frac{k_x x}{L_x}+\frac{k_y y}{L_y})} (\rho(x+1,y,t)+\rho(x-1,y,t)\nonumber\\&+\rho(x,y+1,t)+\rho(x,y-1,t)-4\rho(x,y,t))\nonumber\nonumber\\
  =&D(e^{i2\pi\frac{k_x}{L_x}}+e^{-i2\pi\frac{k_x}{L_x}}
  +e^{i2\pi\frac{k_y}{L_y}}+e^{-i2\pi\frac{k_y}{L_y}}-4)\nonumber\\&
  \sum_{x,y} e^{i2\pi(\frac{k_x x}{L_x}+\frac{k_y y}{L_y})}\rho(x,y,t)\nonumber\\
  =&2D(\cos(\pi\frac{k_x}{L_x})+\cos(2\pi\frac{k_y}{L_y})-2)\nonumber\\&
  \sum_{x,y} e^{i2\pi(\frac{k_x x}{L_x}+\frac{k_y y}{L_y})}\rho(x,y,t).\nonumber\\
\end{align}
This is solved by
\begin{equation}
  \rho(k_x,k_y,t) = \rho(k_x,k_y,0)e^{2D (\cos(2\pi\frac{k_x}{L_x})+\cos(2\pi\frac{k_y}{L_y})-2)t}.
\end{equation}
We expect the same time behavior for the correlation function
\begin{equation}
  \frac{<\rho(k_x,k_y,0)\rho(-k_x,-k_y,t)>}{<\rho(k_x,k_y,0)\rho(-k_x,-k_y,0)>}
  = e^{-\gamma(\mathbf{k}) t}
\label{TimeCorrEqn}
\end{equation}
with
\begin{equation}
\gamma(\mathbf{k})= 2D \left[2-\cos\left(2\pi\frac{k_x}{L_x}\right)+\cos\left(2\pi\frac{k_y}{L_y}\right)\right].
\end{equation}
This behavior is borne out, as is shown for a number of Fourier modes in Fig. \ref{TimeCorr}. We performed additional studies with different relaxation times which all gave similarly satisfactory results.

\begin{figure}
  \includegraphics[width=\columnwidth,clip=true]{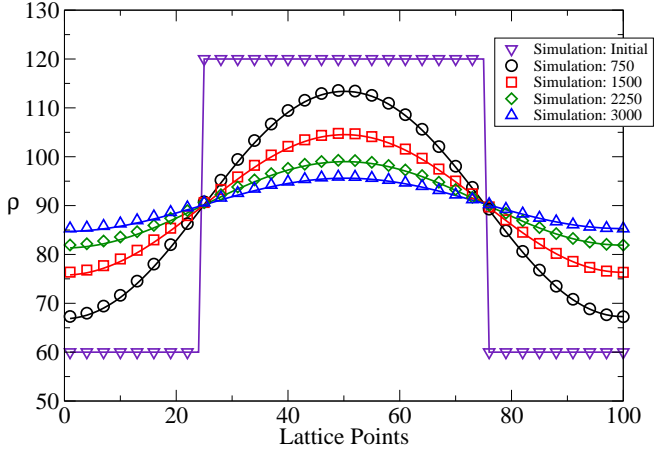}
  \caption{Plot of dynamics of diffusion front with $\rho_1=120$ and $\rho_2=20$. There is good agreement between simulation and theory.} 
  \label{diffusionfront}
\end{figure}

\subsection{Non-equilibrium dynamics}
So far we have only examined the behavior of our system in equilibrium situations. The last test presented in this paper now focuses on an example of non-equilibrium behavior. We study a system that initially represents a step function in the x-direction and which is initially the same in the y-direction. It has a step between two densities $\rho_1$ and $\rho_2$. One can analytically solve the averaged diffusion equation (\ref{DiffAve}) in one dimension on a domain of length $L_x$ with periodic boundary conditions. Then
for a constant diffusion constant $D$ for a block of material with density $\rho_1$ between $0.25 L_x$ and $0.75 L_x$ immersed in material with density $\rho_2$ we obtain the analytic solution
\begin{align}
  \rho^{th}(x,t) = \sum_{i=-\infty}^{\infty}  \frac{\rho_2-\rho_1}{2}&\left[\mbox{erf}\left(\frac{x-(i+0.25)L_x}{2\sqrt{Dt}}\right) \right.\nonumber\\-&\left.\mbox{erf}\left(\frac{x-(i+0.75)L_x}{2\sqrt{Dt}}\right)\right].
\end{align}
We set up a simulation with $L_x=100$ and $L_y=10,000$ and averaged over all y positions. The result of this averaging shows that our numerical solutions are in good agreement with this theoretical solution, as shown in Fig.~\ref{diffusionfront}. 

\begin{figure}
  \subfloat[Global noise \label{vargl}]{\includegraphics[width=\columnwidth,clip=true]{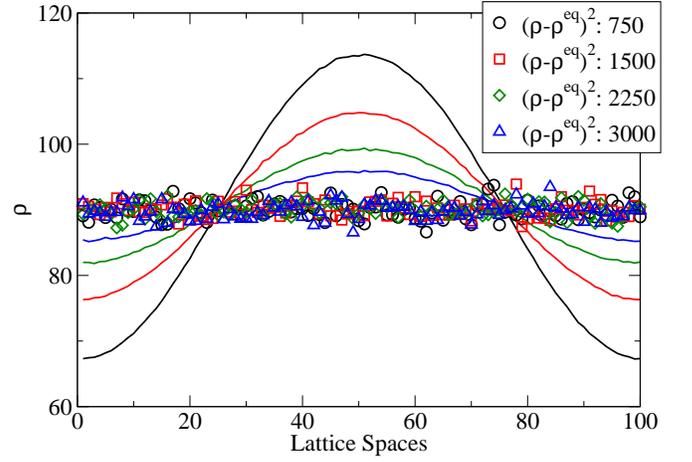}}\\
  \subfloat[Local noise \label{varl}]{\includegraphics[width=\columnwidth,clip=true]{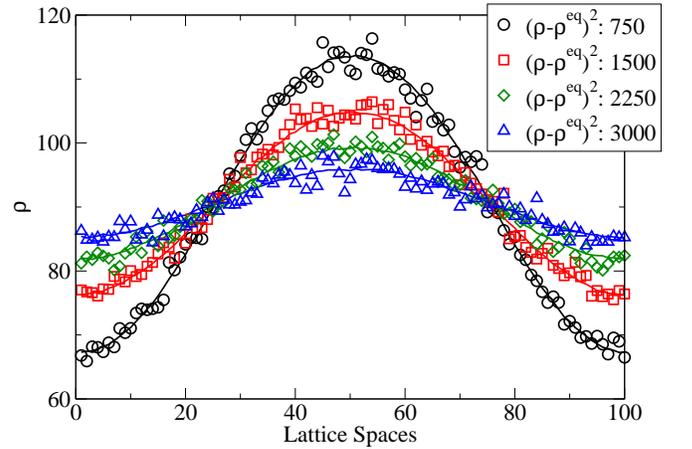}}
  \caption{Plot of variance in $\rho$ in the form $\left(\rho-\rho^{eq}\right)=<\rho>$, with initial density domains of $\rho_1=20$ and $\rho_2=120$. The densities are calculated on a 100x10,000 lattice. The densities are averaged over the y-direction (each x-dimensional lattice space is averaged over 10,000 $y$ values). It is observed that applying noise to $\rho^{eq}$ gives a near constant variance which does not vary in time.}
  \label{varianceplot}
\end{figure}

All this shows, however, is that the effect of the noise averages out as expected. More interestingly we now examined the fluctuations, for which we expect
\begin{equation}
  <(\rho(x,y,t)-\rho^{th}(x,t))^2>=\rho^{th}(x,t).
  \label{rhoth}
\end{equation}
For the global noise implementation, the result is shown in Fig. \ref{vargl}, and the fluctuations are independent of the local density. The key argument for originally proposing local noise implementations given by Kaehler et al. \cite{kaehlerPRE} related to such non-equilibrium situations where the local densities can vary significantly. For our local noise implementation we see that this is indeed borne out. This is shown in Fig. \ref{varl}, where we see excellent agreement with our theoretical prediction of eqn. (\ref{rhoth}).

\section{Conclusion}
We presented the derivation of the simplest possible fluctuating lattice Boltzmann method and, for the first time, were able to obtain a fully consistent derivation. The new derivation allowed us to overcome several difficulties in the derivation of earlier fluctuating lattice Boltzmann methods. The lattice Boltzmann method presented in this paper simulates the fluctuating diffusion equation, but we anticipate that we will be able to use the same approach for deriving a fluctuating lattice Boltzmann method for hydrodynamic systems. This is more complex though, because the equilibrium distribution has more terms, making the calculation of eqns. (\ref{f0f}) and (\ref{f0f0}) more cumbersome.

The consistency of the derivation is reflected in the ability of the new method to recover the expected behavior for all test cases that we have devised. In particular it is remarkable that no deviations of the fluctuation relation for any $k$-modes have been observed. Such deviations are common for other fluctuating methods away from the hydrodynamic regime of small $k$. 

The first derivations of fluctuating lattice Boltzmann methods insisted that the noise amplitude be constant. It was feared that making noise depend on a local density would introduce cumbersome multiplicative noise. Kaehler et al. \cite{kaehlerPRE} suggested using a local noise term on physical grounds, but since the derivation required a transition to Fourier space, this non-constant noise amplitude generated insurmountable difficulties in the analysis. The new analysis presented in this paper allowed us to show that using noise amplitudes that depend on a local density is fully consistent and does not generate difficulties normally found with multiplicative noise. As should be expected the difference between using a fixed noise amplitude and one depending on local noise is most pronounced in non-equilibrium systems with large variations in the density.

A bare-bones implementation of this algorithm was published on GitHub \cite{KyleCodeD2Q5} and is freely available. This implementation only contains the core algorithm, but for ease of readability the various analysis parts of the code have not been included.

In future work we will show how to extend this approach to non-ideal systems, allowing us to apply this method to examine the dynamics in the vicinity of a critical point and in meta stable systems undergoing nucleation. 

\appendix*
\section{\label{App1} Moments of Poisson distributed random variables}
To calculate the expectation value of eqn. (\ref{secondmoment}) we only need to consider two distributions at the same location, so we will drop the spatial dependence here:
\begin{equation}
  <n_i n_j>.
\end{equation}
We need to distinguish the two cases of equal and different values for $i$ and $j$. Let us first consider $ i \neq j$. We then get
\begin{align*}
&<n_i n_j> \\
&= \sum_{n_in_j}P(n_i) P(n_j) n_i n_j\\
&= \exp(-n_i^{eq}-n_j^{eq}) \sum_{n_in_j}\frac{(n_i^{eq})^{n_i}(n_j^{eq})^{n_j}}{n_i!n_j!}n_i n_j\\
&= \exp(-n_i^{eq}-n_j^{eq}) (n_i^{eq}\partial_{n_i^{eq}})(n_j^{eq}\partial_{n_j^{eq}})\sum_{n_in_j}\frac{(n_i^{eq})^{n_i}(n_j^{eq})^{n_j}}{n_i!n_j!}\\
&= \exp(-n_i^{eq}-n_j^{eq}) (n_i^{eq}\partial_{n_i^{eq}})(n_j^{eq}\partial_{n_j^{eq}})\sum_{n n_j}\frac{(n_i^{eq})^{n-n_j}(n_j^{eq})^{n_j}}{(n-n_j)!n_j!}\\
&= \exp(-n_i^{eq}-n_j^{eq}) (n_i^{eq}\partial_{n_i^{eq}})(n_j^{eq}\partial_{n_j^{eq}})\sum_{n}\frac{(n_i^{eq}+n_j^{eq})^n}{n!}\\
&= \exp(-n_i^{eq}-n_j^{eq}) (n_i^{eq}\partial_{n_i^{eq}})(n_j^{eq}\partial_{n_j^{eq}})\exp(n_i^{eq}+n_j^{eq})\\
&=n_i^{eq}n_j^{eq}
\end{align*}
where we used the expression for the binomial formula in line 

Now let us consider the case $ i = j$. We get
\begin{align*}
&<n_i n_i> \\
&= \sum_{n_in_j}P(n_i) n_i n_i\\
&= \exp(-n_i^{eq}) \sum_{n_i}\frac{(n_i^{eq})^{n_i}}{n_i!}n_i n_i\\
&= \exp(-n_i^{eq}) (n_i^{eq}\partial_{n_i^{eq}})(n_i^{eq}\partial_{n_i^{eq}})\sum_{n_in_j}\frac{(n_i^{eq})^{n_i}}{n_i!}\\
&= \exp(-n_i^{eq}) (n_i^{eq}\partial_{n_i^{eq}})[n_i\exp(n_i^{eq})]\\
&=n_i^{eq}n_i^{eq}+n_i^{eq}
\end{align*}
This can be summarized as 
\begin{equation}
<n_i n_j> = n_i^{eq} n_j^{eq} + n_i^{eq} \delta_{ij}.
\label{Poisson2}
\end{equation}
Similarly we get
\begin{align}
  &  <n_in_jn_k> \nonumber\\
  =& n_i^{eq} n_j^{eq} n_k^{eq} + n_i^{eq} n_k^{eq}\delta_{ij} +n_i^{eq} n_j^{eq} \delta_{jk}+n_j^{eq} n_k^{eq} \delta_{ik}\nonumber\\&+n_i^{eq} \delta_{ijk}
\label{f3}
\end{align}

%\begin{acknowledgments}
%\end{acknowledgments}

\bibliography{Fluct}

\end{document}